\documentclass[pre,aps,showpacs]{revtex4}
\usepackage{amssymb}
\usepackage{graphicx}
\begin{document}

\title{Nematic Ordering of Rigid Rods in a Gravitational Field}
\author{Vladimir A. Baulin, Alexei R. Khokhlov}
\affiliation{Physics Department, Moscow State University, Moscow
117234 Russia}
\date{Received 23 December 1998; revised 7 April 1999}

\begin{abstract}
The isotropic-to-nematic transition in an athermal solution of long rigid
rods subject to a gravitational (or centrifugal) field is theoretically
considered in the Onsager approximation. The new feature emerging in the
presence of gravity is a concentration gradient which coupled with the
nematic ordering. For rodlike molecules this effect becomes noticeable at
centrifugal acceleration $g\sim 10^3\div 10^4$ m/s$^2$, while for biological
rodlike objects, such as tobacco mosaic virus, TMV, the effect is important
even for normal gravitational acceleration conditions. Rods are concentrated
near the bottom of the vessel which sometimes leads to gravity induced
nematic ordering. The concentration range corresponding to phase separation
increases with increasing $g$. In the region of phase
separation the local rod concentration, as well as the order parameter,
follow a step function with height.
\end{abstract}


\maketitle

\section{Introduction}

Nematic ordering in a solution of long rigid rods has been studied
theoretically in many papers, starting from the classical papers by Onsager
\cite{1} and Flory \cite{2}. However, there is one aspect of this problem,
which has never been considered, namely that this transition always occurs
in a gravitational field. This field induces a concentration inhomogeneity
within the volume where nematic ordering takes place. Such inhomogeneity
should, in principle, change some of the characteristics of the
liquid-crystalline transition.

The dimensionless parameter associated with gravitational field is $\beta
\equiv mgh/kT$ where $m=m_0-\widetilde{\rho }v$ is the mass of a rod
corrected for buoyancy ($\widetilde{\rho }$ and $v$ are the density of pure
solvent and the volume of the rod, respectively), $h$ is the height, $g$ is
the gravitational acceleration, $T$ is the temperature and $k$ is the
Boltzmann constant. For $h=1$ cm and room temperature conditions, this gives
the following criterion: the inhomogeneity due to the normal gravitational
acceleration ($9.8$ m/s$^2$) becomes important for molecular masses of rod
more than $10^7$ g/mol. Thus, for rigid rods made of common synthetic
macromolecules ($m_0\sim 10^5$ g/mol) this effect can be neglected.

However, in at least two situations the effect of gravity on the problem of
the liquid crystalline transition is important and experimentally relevant.
First, for nematic ordering in solutions of high molecular weight rodlike
biological objects  (such as TMV or virus {\it fd}) \cite{3,4,5,6,7} $m$ can
be very large ($m_0=4\times 10^7$ g/mol for TMV,\cite{4}), and values of  $%
\beta >1$ can easily be reached. Second, instead of normal gravitational
acceleration, one can consider the acceleration in an ultracentrifuge which
can be $10^4\div 10^5$ times larger than ordinary gravity. For at
least these cases, the investigation of the influence of gravitational field
on the nematic ordering in the solution of rigid rods seems to be an
important problem. This problem is solved theoretically below in the Onsager
approximation.

\section{Theoretical model}

The Onsager approach is based on a virial expansion of the free energy of
the solution of rigid rods taking into account steric repulsion only.

Let us choose the Cartesian coordinate system in such a way that the
external field acts in the z direction and put the origin of coordinates at
the bottom of the reservoir (z=0). Further, let us divide the volume of the
vessel V into large number of identical layers aligned perpendicularly to
the field in order that all particles in a given layer has the same
gravitational potential. Here we use the following notation: $dN_z\left(
\Omega \right) =N_zf_z\left( \Omega \right) d\Omega $ is the number of rods
in the layer z with axis directions within the small spatial angle $d\Omega $%
, $N_z$ is the total number of rods in the layer z, and $\zeta =z/h$ denotes
the dimensionless height. Let $f_z\left( \Omega \right) $ be one-particle
orientational distribution function of rods in the layer z. The
normalization for the function $f_z\left( \Omega \right) $ is written in the
familiar form $\int f_z\left( \Omega \right) d\Omega =1$.

With such division of the volume of the vessel into very large number of
layers, the local rod concentration $C^{\prime }(\zeta )$ has the same value
within the layer. Thus, for a given layer one can apply the traditional
Onsager theory \cite{1}, \cite{8,9} justified for homogeneous system. This
theory is valid for dilute enough solutions of very long rods.

In this case, the local free energy of the layer labeled $\zeta$ is written
as
\begin{equation}  \label{eq1}
\frac{\varpi (\zeta )}{kT}=\ln C^{^{\prime }}(\zeta )+\int f\left( \Omega
,\zeta \right) \ln \left( 4\pi f\left( \Omega ,\zeta \right) \right) d\Omega
+C^{^{\prime }}(\zeta )\int f\left( \Omega _1,\zeta \right) f\left( \Omega
_2,\zeta \right) a_2\left( \gamma \right) d\Omega _1d\Omega _2+\int \beta
\zeta f\left( \Omega ,\zeta \right) d\Omega
\end{equation}
where the first term represents the entropy of a translational motion, the
second term is the orientation entropy, the third term describes steric
interaction of rods in the second virial approximation, and the last term is
the average potential energy of a rod in an external gravitational field $%
U_{ext}\left( \zeta \right) /kT=\frac{mgh}{kT}\zeta \equiv \beta \zeta $.

To calculate the third term one assumes that the $\gamma $-dependent second
virial coefficient $a_2$ is the half of excluded volume of two rods \cite{1}%
, thus, $a_2\left( \gamma \right) =L^2D\left| \sin \gamma \right| $, where $%
L $ and $D$ are the length and diameter for long rigid rods and $\gamma $ is
the angle between directions of long axes.

The free energy of the whole system is a sum of free energies of all layers.
If the number of layers is large enough, the sum can be replaced by integral
\begin{equation}  \label{eq2}
\frac F{VkT}=\int_0^1d\zeta C^{^{\prime }}(\zeta )\frac{\varpi (\zeta )}{kT}%
=\int_0^1d\zeta C^{^{\prime }}(\zeta )\left( \ln C^{^{\prime }}(\zeta
)+\sigma (\zeta )+C^{^{\prime }}(\zeta )\rho ^{^{\prime }}(\zeta )+\beta
\zeta \right)
\end{equation}
where

\begin{equation}  \label{eq4}
\sigma (\zeta )\equiv \int f\left( \Omega ,\zeta \right) \ln \left( 4\pi
f\left( \Omega ,\zeta \right) \right) d\Omega
\end{equation}
is the orientational entropy of the layer $\zeta $ and

\begin{equation}  \label{eq5}
\rho ^{\prime }(\zeta )=\int f\left( \Omega _1,\zeta \right) f\left( \Omega
_2,\zeta \right) a_2\left( \gamma \right) d\Omega _1d\Omega _2
\end{equation}
is the second virial coefficient of interaction of two rods.

To obtain the equilibrium distribution function we should take into account
the possibility of the formation of a phase boundary between the nematic
phase at the bottom of the vessel and isotropic phase on top. We denote the
height of the boundary position in the vessel as $\zeta =x$, so the volumes
occupied by the nematic and isotropic phases are $V_a=xV$ and $V_i=\left(
1-x\right) V$, respectively. With this, the free energy of the whole system
eq. (\ref{eq2}) becomes

\begin{equation}  \label{eq22}
\frac F{VkT}=\int_0^xd\zeta C_a^{^{\prime }}(\zeta )\frac{\varpi _a(\zeta )}{%
kT}+\int_x^1d\zeta C_i^{^{\prime }}(\zeta )\frac{\varpi _i(\zeta )}{kT},
\end{equation}
where

\begin{equation}  \label{eq22a}
\frac{\varpi _a(\zeta )}{kT}=\ln C_a^{^{\prime }}(\zeta )+\sigma _a(\zeta
)+C_a^{^{\prime }}(\zeta )\rho _a^{^{\prime }}(\zeta )+\beta \zeta
\end{equation}
is the local free energy of the nematic phase, and

\begin{equation}  \label{eq22i}
\frac{\varpi _i(\zeta )}{kT}=\ln C_i^{^{\prime }}(\zeta )+\sigma _i(\zeta
)+C_i^{^{\prime }}(\zeta )\rho _i^{^{\prime }}(\zeta )+\beta \zeta
\end{equation}
is the local free energy of the isotropic phase.

To calculate the equilibrium distribution function $f\left( \Omega ,\zeta
\right) $ one should minimize the functional (\ref{eq22}) with respect to
this function. The direct minimization of functional (\ref{eq22}) leads to a
nonlinear integral equation, which can be solved only numerically \cite
{10,11}. In the case where the volume of the vessel consists of two phases
separated by a phase boundary, one should realize that $f\left( \Omega
,\zeta \right) $ follows a step function with the variation of $\zeta $ with
the function $f_a(\Omega ,\zeta )$ in the anisotropic part significantly
different from that in the isotropic phase $f_i(\zeta )=1/4\pi $. To
evaluate the distribution function $f_a\left( \Omega ,\zeta \right) $ in the
nematic phase we apply an approximate variational method with a trial
function depending on variational parameter $\alpha $.

Substituting this function in eqs. (\ref{eq22}), (\ref{eq22a}), (\ref{eq22i}%
) and minimizing with respect to $\alpha (\zeta )$, we have the following
equation for definition of variational function $\alpha (\zeta )$
\begin{equation}  \label{eq6}
\frac{d\sigma _a(\alpha )}{d\alpha }+C_a^{^{\prime }}(\zeta )\frac{d\rho
_a^{^{\prime }}(\alpha )}{d\alpha }=0
\end{equation}

However, the trial function proposed by Onsager \cite{1} $f(\Omega ,\alpha
)=\left( \alpha /(4\pi \sinh \Theta )\right) \cosh \left( \alpha \cos \Theta
\right) $, where $\Theta $ is the angular deviation of a rod from the
director, still leads to rather complicated integral equation. Therefore,
following \cite{9} for the sake of simplicity we used a trial function of
simpler form
\begin{equation}  \label{eq7}
f_a\left( \Omega ,\alpha (\zeta )\right) =\frac{\alpha (\zeta )}{4\pi }%
\left\{
\begin{array}{c}
e^{- \frac{\alpha (\zeta )\Theta ^2}2}\text{ },0<\Theta <\frac \pi 2 \\
e^{-\frac{\alpha (\zeta )(\pi -\Theta )^2}2},\frac \pi 2<\Theta <\pi
\end{array}
\right.
\end{equation}
with an approximate normalization (precise up to terms of order $O\left(
e^{-\alpha }\right) $).

This trial function is suitable for approximate evaluation of $\sigma
_a(\alpha (\zeta ))$ in case of highly ordered state \cite{9}
\begin{equation}  \label{eq8}
\sigma _a\left( \alpha (\zeta )\right) \equiv \int f_a\left( \Omega ,\alpha
(\zeta )\right) \ln \left( 4\pi f_a\left( \Omega ,\alpha (\zeta )\right)
\right) d\Omega \approx \ln \alpha (\zeta )-1,
\end{equation}
and the dimensionless second virial coefficient in the anisotropic phase
\cite{9}
\begin{equation}  \label{eq9}
\rho _a(\zeta )=\frac{\rho _a^{^{\prime }}(\zeta )}b=\frac 4\pi \int
f_a\left( \Omega _1,\alpha (\zeta )\right) f_a\left( \Omega _2,\alpha (\zeta
)\right) \left| \sin \gamma \right| d\Omega _1d\Omega _2\approx \frac 4{%
\sqrt{\pi \alpha (\zeta )}}
\end{equation}
with notation $b\equiv \left\langle \left\langle \rho ^{^{\prime }}(\zeta
)\right\rangle _i\right\rangle _i=L^2D\left\langle \left\langle \left| \sin
\gamma \right| \right\rangle _i\right\rangle _i=\frac \pi 4L^2D$; the value
of $b$ being equal to half of average excluded volume of two arbitrary
oriented rods.

The corresponding expressions in the isotropic phase are
\begin{equation}
\sigma _i=0\text{ and }\rho _i=1\text{.}  \label{eq10}
\end{equation}

In the above formulas the index $i$ and $a$ refers to the isotropic phase
and the nematic phase respectively, and the angular brackets designate the
average with respect to the isotropic distribution function $f_i\left(
\Omega \right) =1/4\pi $.

Substituting the calculated values back in eq. (\ref{eq6}), yields the
following expression for the function $\alpha (\zeta )$:
\begin{equation}
\sqrt{\alpha (\zeta )}=\frac{2C_a(\zeta )}{\sqrt{\pi }},  \label{eq11}
\end{equation}
where $C_a(\zeta )=C_a^{^{\prime }}(\zeta )b$ is the dimensionless local rod
concentration in the nematic phase.

Thus, eqs. (\ref{eq22a}), (\ref{eq22i}) and (\ref{eq10}) , (\ref{eq11}) give

\begin{equation}  \label{eq12a}
\frac{\varpi _a(\zeta )}{kT}=\ln C_a^{^{\prime }}(\zeta )+2\ln \frac{%
2C_a(\zeta )}{\sqrt{\pi }}+1+\beta \zeta
\end{equation}
and

\begin{equation}  \label{eq12i}
\frac{\varpi _i(\zeta )}{kT}=\ln C_i^{^{\prime }}(\zeta )+C_i(\zeta )+\beta
\zeta .
\end{equation}

The chemical potentials in the phases can be also obtained

\begin{equation}  \label{eq13}
\mu _{a,i}=\varpi _{a,i}+p_{a,i}(\zeta )v_{a,i}(\zeta ),
\end{equation}
where $v_{a,i}(\zeta )=1/C_{a,i}^{^{\prime }}(\zeta )$ is the local specific
volume and $p_{a,i}(\zeta )$ is the pressure in the layer $\zeta $:

\begin{equation}  \label{eqp}
p_{a,i}(\zeta )=-\frac{\partial \varpi _{a,i}(\zeta )}{\partial
v_{a,i}(\zeta )}=\left( C_{a,i}^{^{\prime }}(\zeta )\right) ^2\frac{\partial
\varpi _{a,i}(\zeta )}{\partial C_{a,i}^{^{\prime }}(\zeta )}
\end{equation}

The calculation of the pressure in the nematic and isotropic phases gives
(compare with ref. \cite{9})

\begin{equation}  \label{eqpa}
p_a(\zeta )=3C_a^{^{\prime }}(\zeta ),
\end{equation}

\begin{equation}  \label{eqpi}
p_i(\zeta )=C_i^{^{\prime }}(\zeta )\left( 1+C_i(\zeta )\right) .
\end{equation}

Consequently, eqs. (\ref{eq13}), (\ref{eq12a}), (\ref{eq12i}) and (\ref{eqp}%
) give

\begin{equation}  \label{eqma}
\frac{\mu _a(\zeta )}{kT}=\ln C_a^{^{\prime }}(\zeta )+2\ln \frac{2C_a(\zeta
)}{\sqrt{\pi }}+4+\beta \zeta
\end{equation}
and

\begin{equation}  \label{eqmi}
\frac{\mu _i(\zeta )}{kT}=\ln C_i^{^{\prime }}(\zeta )+2C_i(\zeta )+1+\beta
\zeta
\end{equation}

In the equilibrium the chemical potential is independent on height and is
the same in the both phases, thus one can obtain the equilibrium local
concentrations in the phases:

\begin{equation}  \label{eqcona}
C_a^{^{\prime }}(\zeta )=\frac{C_a^{^{\prime }}}{I_1(x)}e^{-\frac{\beta
\zeta }3},
\end{equation}
where $C_a^{^{\prime }}$ is the average concentration in the nematic phase, $%
I_1(x)\equiv \int_0^xe^{-\frac{\beta \zeta }3}d\zeta $ is the normalization
factor.

Also

\begin{equation}  \label{eqconi}
C_i^{^{\prime }}(\zeta )=\frac 1{2b}LW\left( 2be^{-\beta \zeta -1+\mu
_i}\right) ,
\end{equation}
where the function $LW\left( x\right) $ corresponds to a solution of
equation $LW\exp (LW)=x$. For dilute solutions we can use the simple
asymptotic form of this special function: $LW\left( x\right) =x+O(x^2)$.

Thus,
\begin{equation}  \label{eqconi2}
C_i^{^{\prime }}(\zeta )\approx \frac{C_i^{^{\prime }}}{I_2(x)}e^{-\beta
\zeta },
\end{equation}
with $I_2(x)\equiv \int_x^1e^{-\beta \zeta }d\zeta $.

The equilibrium concentrations in the phases are determined by the following
coexistence relations at the boundary

\begin{equation}  \label{eql}
\left\{
\begin{array}{c}
p_a(x)=p_i(x), \\
\mu _a(x)=\mu _i(x),
\end{array}
\right.
\end{equation}

Substituting calculated values of chemical potentials (\ref{eqma}), (\ref
{eqmi}) and pressures (\ref{eqpa}), (\ref{eqpi}) with obtained
concentrations in the phases (\ref{eqcona}), (\ref{eqconi}) the coexistence
relations (\ref{eql}) is written as

\begin{equation}  \label{eqlas}
\left\{
\begin{array}{c}
3 \frac{C_a^{^{\prime }}}{I_1(x)}e^{-\frac{\beta x}3}=\frac{C_i^{^{\prime }}%
}{I_2(x)}e^{-\beta x}\left( 1+\frac{C_i}{I_2(x)}e^{-\beta x}\right) , \\
\ln \frac{C_a^{^{\prime }}}{I_1(x)}+2\ln \frac{2C_a}{I_1(x)\sqrt{\pi }}%
+3=\ln \frac{C_i^{^{\prime }}}{I_2(x)}+2\frac{C_i}{I_2(x)}e^{-\beta x},
\end{array}
\right.
\end{equation}
where $C_a\equiv C_a^{^{\prime }}b$ and $C_i\equiv C_i^{^{\prime }}b$ are
dimensionless average concentrations in the nematic and isotropic phases,
respectively.

If the gravitational field is absent ($g\rightarrow 0$), eqs. (\ref{eqlas})
are reduced to corresponding equations for homogeneous system \cite{9}.

\section{Obtained results}

\begin{figure}
\includegraphics[height=6cm,width=8cm]{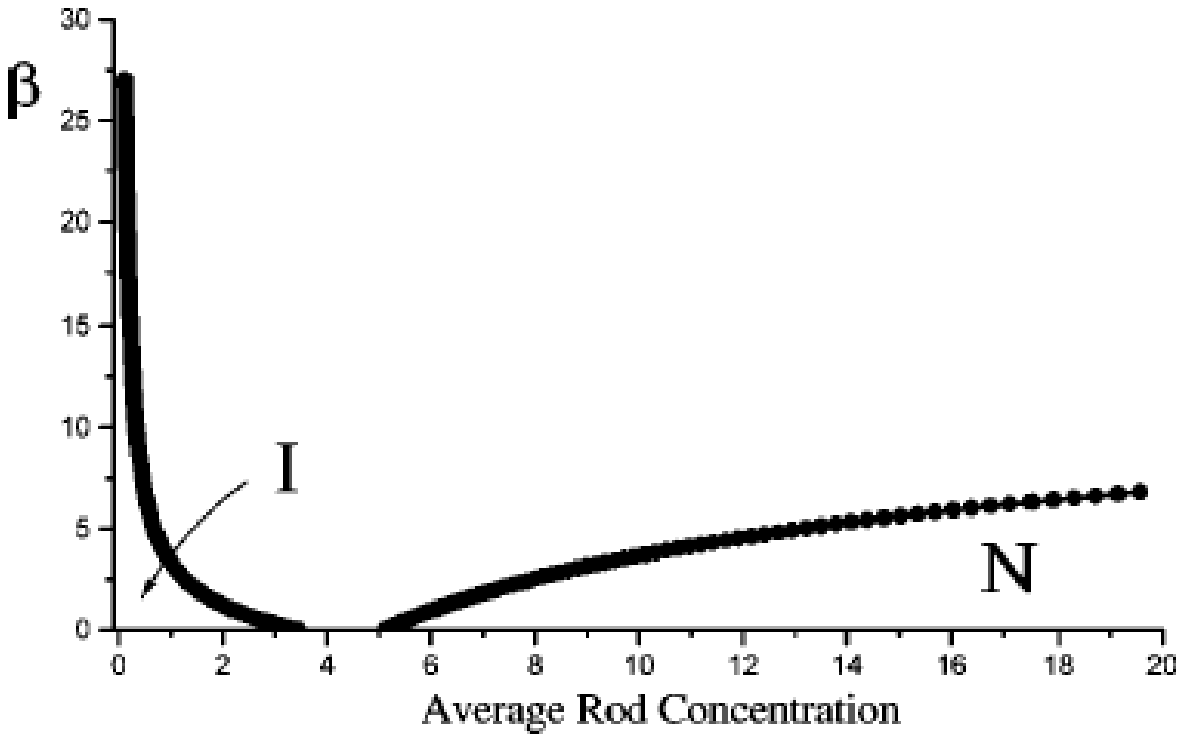}
\caption{ Phase diagram in the variables $\beta$  -- $C$ for
athermal solution of long rigid rods. Here $\beta\equiv mgh/kT$ is
the dimensionless parameter associated with external field, $C$ is
the dimensionless average rod concentration. Label $N$ designates
the nematic phase, $I$ the isotropic phase while $N+I$ corresponds
to phase separation region.} \label{fig1}
\end{figure}

The numerical solution of eqs. (\ref{eqlas}) gives the values of average
dimensionless concentrations of the nematic and isotropic phases, $C_a$ and $%
C_i$, coexisting at equilibrium. The phase diagram in the variables average
rod concentration $C$ -- dimensionless parameter $\beta $ is shown in Fig.\
\ref{fig1}. This diagram has three main regions. In the region labeled by
letter $I$, entire solution of rods is isotropic (corresponding height of
the boundary $x=0$). In the region between the curves of coexistence $%
C_i(\beta )$ and $C_a(\beta )$, the solution is separated in the isotropic
and nematic phases with an interphase boundary between them ($0<x<1$). In
the region designated by letter $N$, the entire volume of the vessel is
occupied by the nematic phase ($x=1$).

This diagram reveals that gravity facilitates formation of the nematic phase
(at least at the bottom of the vessel) and the region of phase separation
becomes very broad even for rather low values of $\beta $.

\begin{figure}
\includegraphics[height=6cm,width=8cm]{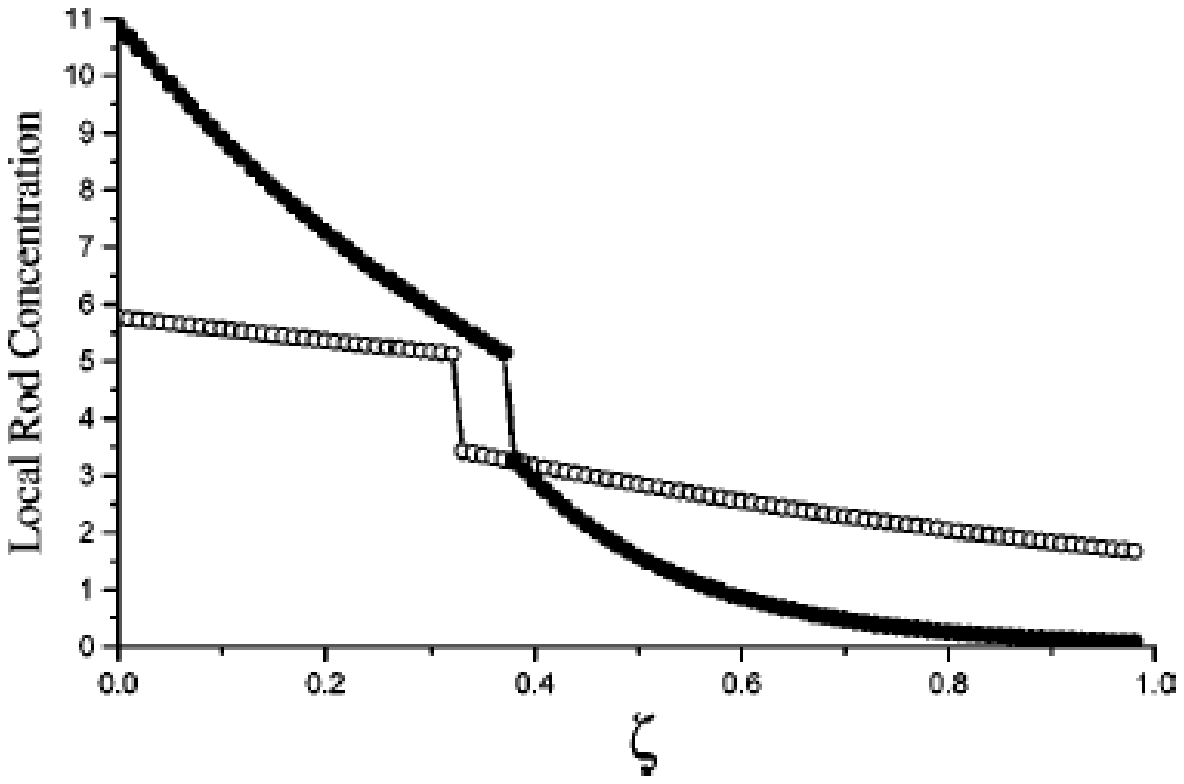}
\caption{ Dependence of the dimensionsless local rod concentration
$C(\zeta ) $ on the dimensionless height $\zeta $ at fixed value
of $\beta$. Open circles correspond to $\beta=1.1$; solid circles
correspond to $\beta=6.1$; dimensionless total rod concentration
is $C=3.4$.} \label{fig2}
\end{figure}

Dependence of the local rod concentration on the height $\zeta $ at fixed
value of $\beta $ is shown in Fig. \ref{fig2}. The concentrations $C_a(\zeta
)$ and $C_i(\zeta )$ obey the barometric distribution according to eqs. (\ref
{eqcona}) and (\ref{eqconi2}), respectively. The concentrations of the
nematic and isotropic phases in the boundary layer coincide with that in the
absence of the field ($C_a(x)=5.12$ and $C_a(x)=3.45$ \cite{9}). This is the
case because all the rods within a given layer exhibit the same
gravitational potential. Thus, the rod concentration follows a step function
with a jump at the phase boundary.

\begin{figure}
\includegraphics[height=6cm,width=8cm]{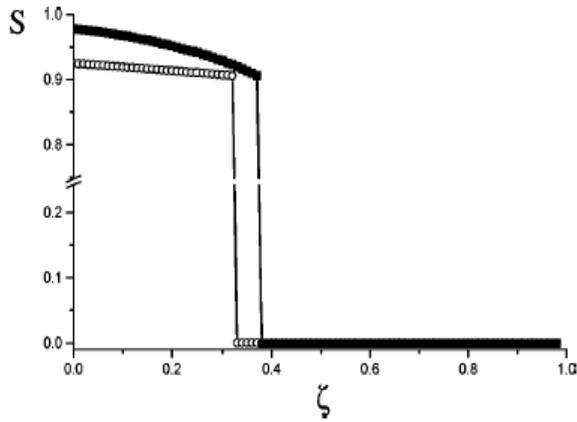}
\caption{ Dependence of the order parameter $S$ on the dimensionless height $%
\zeta $ at fixed values of $\beta$. Open circles correspond to $\beta=1.1$;
solid circles correspond to $\beta=6.1$; dimensionless total rod
concentration is $C=3.4$.}
\label{fig3}
\end{figure}

The corresponding change with height of the order parameter $S(\zeta )=\int
P_2\left( \cos \alpha (\Omega ,\zeta )\right) f(\Omega ,\zeta )d\Omega $ is
shown in Fig. \ref{fig3}. The order parameter $S(\zeta )$ in the nematic
phase is overstated due to the approximate trial function of the form (\ref
{eq7}). With increase of $\zeta $ the order parameter decreases to the value
corresponding to a nematic phase coexisting with isotropic one in the
absence of gravity and then falls to zero.

The Onsager approximation (\ref{eq1}) used in this paper is valid for low
rod concentrations (volume fraction $\varphi \leq 0.1$). However, with
increase of $\beta $ the local rod concentration at the bottom of the vessel
gradually increases. Thus, for high values of $\beta $ barometric
distribution (\ref{eqcona}) is no longer valid.

To generalize the Onsager theory for the case of high rod concentrations one
can use the Parsons approximation\cite{12}, which aims to improve the second
virial coefficient (\ref{eq9}) by means of additional multiplier depending
on mole fraction of rods, as well as some others generalizations (cf. refs.
\cite{13,14,15,16}). Nevertheless, the calculations with nonanalitic
distribution arising from such an approach are rather complicated and lead
to additional integral equation. In most practical cases, except
sedimentation in ultracentrifuge, the values of $\beta $ are not too high
(e.g. for TMV $\beta $ is slightly above the unity), and traditional second
virial approximation is quite justified.

\begin{figure}
\includegraphics[height=6cm,width=8cm]{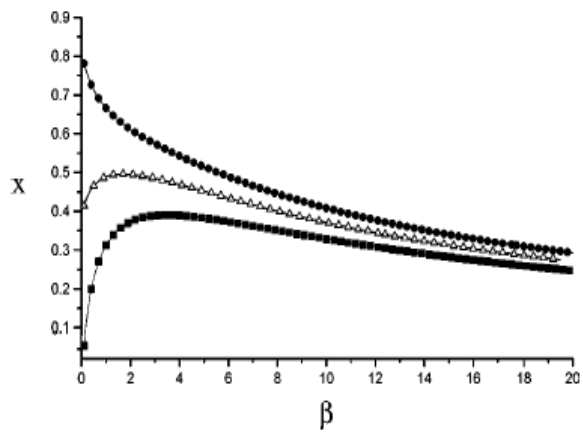}
\caption{ The phase boundary position plotted as a function of
$\beta$ for different values of dimensionless total rod
concentration $C$. The solid circles refer to $C=4.8$; the open
triangles to $C=4.1$; the solid squares to $C=3.4$.} \label{fig4}
\end{figure}

The position of the phase boundary vs. $\beta $ for different values of
total rod concentration is shown in Fig. \ref{fig4}. These plots lead to the
following conclusions. If the total rod concentration is low enough (i.e.
the greater part of the vessel is occupied by the isotropic phase, $C\sim C_i
$), the increasing gravity induces the isotropic-to-nematic transition and
phase boundary shifts toward the top of the vessel. This process is observed
until $\beta \sim 1$ and then the shift of the boundary stops, and the
volume of nematic phase even slightly decreases (bottom phase is becoming
denser under gravity; solid squares). If the total rod concentration is high
enough ($C\sim C_a$) the nematic phase simply shrinks under gravity starting
from the top of the vessel and the position of the phase boundary becomes
gradually lower (solid circles).

Furthermore, it is noteworthy to emphasize the important conclusion arising
from the form of Fig. \ref{fig1}. The right branch of the plot rises very
rapidly as $\beta $ gets large, thus, remaining within the framework of
general concepts dealing with spatially homogeneous phases one could suggest
that the concentration of rods in the nematic phase should also rapidly
increase with $\beta $. However, the increase in the average concentration
in the nematic phase is not as drastic as it follows from Fig. \ref{fig1}.
The general reason is that the ''rule of lever'' cannot be applied for the
present system, because we are dealing with spatially inhomogeneous phases.

\begin{figure}
\includegraphics[height=6cm,width=8cm]{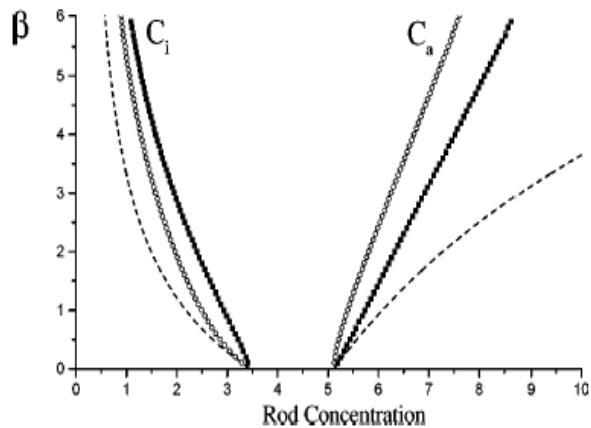}
\caption{ Average concentrations of rods in isotropic and nematic
phases as a function of $\beta $ for different values of total rod
concentration $C$. Open circles correspond to $C=3.4$, solid
squares correspond to $C=4.8$.} \label{fig5}
\end{figure}

The average concentrations of nematic and isotropic phases, corresponding to
phase separation, are shown in Fig. \ref{fig5}. This plot demonstrates that
the average concentrations in both phases at fixed value of $\beta $ do
depend on the total concentration of rods and they do {\it not coincide}
with the concentrations corresponding to the curves of Fig. \ref{fig1}
(dashed lines in Fig. \ref{fig5}). This is because $\beta \sim h$, where $h$
is the total height of the vessel. Thus, the parameter $\beta $ is different
for separate phases and for the system as a whole. That is why the average
concentrations of the phases lie within the region of phase separation shown
in Fig. \ref{fig1}.

\section{Conclusions}

Gravitational or centrifugal external fields facilitate liquid-crystalline
transition at the bottom of the vessel and broadens the region of phase
separation. This phenomenon should be noticeable for biological rod-like
objects or common lyotropic molecules sedimenting in a centrifugal field.
This seems to be an important problem which requires experimental
investigation.

\acknowledgments The authors thank Dr. S. Fraden who has drawn
their attention to this unsolved problem.

\newpage

\end{document}